\documentclass[12pt]{article}

\usepackage{amsmath,amssymb,amsbsy,amsfonts,amsthm,latexsym,
                        amsopn,amstext,amsxtra,euscript,amscd,color}
\usepackage{color,xcolor}
\usepackage{latexsym}
\usepackage{cite}
\usepackage{amsfonts}
\usepackage{amsfonts,mathrsfs}
\usepackage{graphicx}
\usepackage{psfrag}
\usepackage{subfigure}
\usepackage{url}
\usepackage{stfloats}
\usepackage{amsmath}
\usepackage{algorithm}
\usepackage{algorithmic}
\usepackage{ulem}

\usepackage{hyperref}
\hypersetup{colorlinks=true}

\newtheorem{theorem}{Theorem}
\newtheorem{lemma}{Lemma}

\newtheorem{conjecture}{Conjecture}

\newcommand{\quash}[1]{}
\begin{document}

\title{Arithmetic Autocorrelation of Binary $m$-Sequences}

\author{Zhixiong Chen$^{1}$, Zhihua Niu$^{1,2}$, Yuqi Sang$^{2}$ and Chenhuang Wu$^{1}$\\ \\
1. Key Laboratory of Applied Mathematics of Fujian Province University,\\
 Putian University, Putian, Fujian 351100, P. R. China\\
2. School of Computer Engineering and Science,\\
 Shanghai University, Shanghai 200444, P. R. China
}

\maketitle

\begin{abstract}
An $m$-sequence is the one of the largest period among those produced by a linear feedback shift register. It possesses several desirable features of pseudorandomness
such as balance, uniform pattern distribution and ideal autocorrelation for applications to communications.
However, it also possesses undesirable features such as low linear complexity.
Here we prove a nontrivial upper bound on its arithmetic autocorrelation, another figure of merit introduced by Mandelbaum for error-correcting codes
and later investigated by Goresky and Klapper for FCSRs. The upper bound is close to half of the period and hence rather large, which gives an undesirable feature.
\end{abstract}



\section{Introduction}

\indent

An $m$-sequence is the one of longest length produced by a linear feedback shift register (LFSR, for short), that is, its period is $2^n-1$ if it is generated by an $n$-order LFSR.
It has a wide range of applications in the field of communications such as spread spectrum communications and code division multiple access (CDMA).

Any binary $m$-sequence $\mathcal{M}=(m_i)_{i\geq 0}$ over the finite field $\mathbb{F}_2=\{0,1\}$ of period $T=2^n-1$ has many desirable features but also some undesirable ones. For example, it
\begin{itemize}
\item is balanced and has uniform run property, see \cite[Chapter 5]{GG}.

\item has ideal (classical) autocorrelation, i.e., $\sum\limits_{0\leq i<T} (-1)^{m_i\oplus m_{i+\tau}}\in \{-1, T\}$ for $0\leq \tau<T$,
where $\oplus$ is the addition modulo 2,  see \cite[Chapter 5]{GG}.

\item  has maximum 2-adic complexity, which is $\log_2(2^T-1)$, see \cite{TQ2010}.

\item has, however, low linear complexity, which is $n$. So it cannot resist the Berlekamp-Massey algorithm \cite{M1969}.

\item has the ``shift-and-add property": for each $0\leq \tau<T$, there is a corresponding $\tau'$ such that $m_i\oplus m_{i+\tau}=m_{i+\tau'}$ for $i\geq 0$ \cite[Theorem 5.3]{GG}.
So it is very vulnerable for the fast correlation attack, see \cite{ALHJ2012,MS1989}.

\end{itemize}

Indeed, there is a huge literature about the study of $m$-sequences, which might involve almost all pseudorandom measures for sequences, we refer the reader to (but not limited to) \cite{D2021,GG,H1999}.

In this work, we will discuss the arithmetic autocorrelation introduced by Mandelbaum \cite{M1967} (for binary $m$-sequences).
In the \emph{arithmetic autocorrelation}, a sequence is added to a shift of itself
with carry, rather than bit by bit modulo 2. More precisely, the 2-adic numbers are employed for defining the arithmetic correlation \cite{GK1997}.
Let $\mathcal{S}=(s_i)_{i\geq 0}$ be a $T$-periodic\footnote{For our purpose, we only restrict periodic sequences.
For arithmetic autocorrelation of non-periodic sequences the reader is referred to \cite{GK2008,GK2012}.} binary sequence over $\mathbb{F}_2$ and
$\mathcal{S}^{(\tau)}=(s_{i+\tau})_{i\geq 0}$ the $\tau$-shift of $\mathcal{S}$. Write
$$
\underline{s}=\sum\limits_{i=0}^{\infty}s_i2^i, ~~~ \underline{s}^{(\tau)}=\sum\limits_{i=\tau}^{\infty}s_i2^{i-\tau},
$$
which are the 2-adic numbers uniquely associated with $\mathcal{S}$ and $\mathcal{S}^{(\tau)}$, respectively.
From $\underline{s}-\underline{s}^{(\tau)}$, we recover a (unique) binary sequence $\mathcal{U}=(u_i)_{i\geq 0}$ with
$\underline{s}-\underline{s}^{(\tau)}=\sum\limits_{i=0}^{\infty}u_i2^{i}$, which is eventually periodic over $\mathbb{F}_2$.
Then the arithmetic autocorrelation of $\mathcal{S}$, denoted by $\mathcal{A}^{A}_{\mathcal{S}}(\tau)$, is
$$
\mathcal{A}^{A}_{\mathcal{S}}(\tau)=\sum\limits_{i=\jmath}^{\jmath+T-1}(-1)^{u_i},
$$
where $\mathcal{U}$ is strictly periodic from $u_{\jmath}$ on. So we need to determine the number of ones (or zeros) in one period of $\mathcal{U}$.

It is desirable that the absolute values of the arithmetic autocorrelations are as small as possible.
 A sequence, such as $\ell$-sequence \cite{GK2008}, has ideal arithmetic autocorrelations, in the sense that
they are zero for nontrivial shifts.
In particular, Goresky and Klapper \cite{GK1997} stated a potential application (of arithmetic correlation)
for simultaneous synchronization and identification in a multiuser environment.
The reader is referred to the works by
Goresky and Klapper \cite{GK1997,GK2008,GK2011} or the monograph \cite{GK2012} for more background and results on arithmetic auto/cross-correlations.

Very recently, Hofer, Merai and Winterhof  \cite{HW2017,HMW2017} studied  the arithmetic autocorrelation for some special binary sequences,
including the Legendre sequences and the Sidelnikov-Lempel-Cohn-Eastman sequences. In particular, they developed a relation between
the arithmetic autocorrelation and the higher-order (periodic) correlation measure, see \cite[Theorem 1]{HMW2017} for details. As a consequence,
they stated in \cite[Sect.4]{HMW2017} that for a truly random sequence of
length/period $T$, its arithmetic autocorrelation is at most of
order of magnitude $T^{3/4}(\log_2 T)^{1/2}$. However, \cite[Theorem 1]{HMW2017} fails for $m$-sequences, since  the maximum  3-order (periodic) correlation measure (see the notion in \cite{HMW2017}) of an $m$-sequence
is equal to the period due to the ``shift-and-add property". That is, \cite[Theorem 1]{HMW2017} gives nothing about the arithmetic autocorrelation of $m$-sequences.

Fortunately, we find that the `binary subtraction' (see below for the notion) of two binary vectors can help us to estimate
the arithmetic autocorrelation for binary $m$-sequences, which is our main contribution in this letter stated in the following theorem.

\begin{theorem}\label{thm-AC}
Let $\mathcal{M}$ be a binary $m$-sequence over $\mathbb{F}_2$ of period $2^n-1$ for some $n\in \mathbb{N}$. Then the arithmetic autocorrelation of $\mathcal{M}$ satisfies
$$
\left| \mathcal{A}^{A}_{\mathcal{M}}(\tau) \right|\leq  2^{n-1}-1, ~~~\mathrm{where}~~~ 1\leq \tau< 2^n-1.
$$
\end{theorem}
Certain experimental tests tell us that the upper bound is achievable. For the proof, we present some ideas in Sect.2.
In Sect.3, we develop many experimental tests, which back the theorem up
and from which a conjecture arises. We also include some interesting problems for further study.

\section{Ideas for estimating the arithmetic autocorrelation of $m$-sequences}

\subsection{Preparations}
\indent

According to  \cite[Proposition 2]{GK1997}, Hofer and Winterhof \cite[PP.238-239]{HW2017} re-established the process of computing the arithmetic correlation for periodic sequences.
It is easy understanding and hence we demonstrate the computation as follows:

For a $T$-periodic binary sequence $\mathcal{S}=(s_i)_{i\geq 0}$ over $\mathbb{F}_2$ and
its  $\tau$-shift $\mathcal{S}^{(\tau)}=(s_{i+\tau})_{i\geq 0}$, write
$$
S(X)= \sum\limits_{0\leq i<T}s_iX^{i}, ~~~ S^{(\tau)}(X)= \sum\limits_{0\leq i<T}s_{i+\tau}X^{i}.
$$
We compute the integer $S(2)-S^{(\tau)}(2)$ in $\mathbb{Z}$ and represent the absolute value $|S(2)-S^{(\tau)}(2)|$ as a binary expansion:
$$
|S(2)-S^{(\tau)}(2)|= \sum\limits_{0\leq i<T} w_i 2^i, ~~~ w_i\in\{0,1\}.
$$
Let $N_0=|\{i : w_i=0, 0\leq i<T\}|$ and $N_1=|\{i : w_i=1, 0\leq i<T\}|$. Then we compute the arithmetic autocorrelation of $\mathcal{S}$:
\begin{equation}\label{AA+}
\mathcal{A}^{A}_{\mathcal{S}}(\tau)=N_0-N_1=T-2N_1
\end{equation}
if $S(2)>S^{(\tau)}(2)$, and otherwise
\begin{equation}\label{AA-}
\mathcal{A}^{A}_{\mathcal{S}}(\tau)=N_1-N_0=2N_1-T.
\end{equation}

For example, let $\mathcal{S}$ be 15-periodic with the first period $0001\uwave{1}1101011001$. If we choose $\tau=4$, the 4-shift is
$\uwave{1}11010110010001$. We have
$$
S(X)=X^3+X^4+ X^5+ X^6+X^8+ X^{10}+X^{11}+ X^{14}
$$
and
$$
S^{(4)}(X)=1+X+X^2+ X^4+ X^6+X^7+ X^{10}+ X^{14}.
$$
Then we compute
$$
S(2) - S^{(4)}(2)=19832-17623=2209=1+2^5+2^7+2^{11},
$$
from which we see that $N_1=4$ and $N_0=15-N_1=11$. Hence we get $\mathcal{A}^{A}_{\mathcal{S}}(4)=N_0-N_1=7$ by Eq.(\ref{AA+}).
For other $\tau$ one can deal with the calculations in a similar way.\\

We find that there are two operations in the computation above, one is the integer substraction, the other is the binary expansion of an integer.
For our purpose, we write the first period of $\mathcal{S}=(s_i)_{i\geq 0}$ of period $T$ as a $T$-dimensional binary vector $(s_0,s_1,\ldots,s_{T-1})$ and
introduce the `binary subtraction' for two vectors to get a new vector with entries in $\{\overline{1},0,1\}$, where $\overline{1}=-1$, according to
\begin{equation}\label{S-S}
S(2)-S^{(\tau)}(2)= \sum\limits_{0\leq i<T} (s_i-s_{i+\tau}) 2^i, ~~~ s_i-s_{i+\tau}\in \{\overline{1},0,1\}.
\end{equation}
We need to determine the number of 1's and $\overline{1}$'s in $s_i-s_{i+\tau}\in \{\overline{1},0,1\}$ for $0\leq i<T$. This is why we want to consider the binary subtraction.

So in general, we denote by $\boxminus$ the \emph{binary subtraction} of $\{0,1\}$ as follows:
$$
0\boxminus 0=0, ~~ 1\boxminus 0=1, ~~ 1\boxminus 1=0, ~~ 0\boxminus 1=\overline{1}.
$$
We note that, unlike the general subtraction for integers, $\boxminus$ is done without carry.
For two $T$-dimensional vectors $\alpha=(a_0,a_1,\ldots,a_{T-1})$ and $\beta=(b_0,b_1,\ldots,b_{T-1})$ in $\{0,1\}^{T}$,
we have
$$
\alpha \boxminus \beta =(a_0 \boxminus b_0, a_1 \boxminus b_1, \ldots, a_{T-1} \boxminus b_{T-1})\in \{\overline{1},0,1\}^T
$$
In fact, this is similar to the binary addition $\oplus$ (without carry), the addition modulo 2, we have $\alpha \oplus \beta=(a_0\oplus b_0,a_1\oplus b_1,\ldots,a_{T-1}\oplus b_{T-1})\in \{0,1\}^T$.
As usual, the \emph{weight}, denoted by $wt(\cdot)$, of a vector is the number of non-zeros in the vector. Then $wt(\alpha \boxminus \beta)$ equals to
the total number of 1's and $\overline{1}$'s. We have the following simple statement.

\begin{lemma}\label{lemma-boxminus}
Let $\alpha,\beta\in \{0,1\}^{T}$ be two $T$-dimensional binary vectors.

(i). We have
$$
wt(\alpha \boxminus \beta)=wt(\alpha \oplus \beta).
$$

(ii). Let $N_x$ be the number of $x\in \{\overline{1},0,1\}$ in $\alpha \boxminus \beta$. We have
$$
N_1=\frac{wt(\alpha \boxminus \beta)+wt(\alpha)-wt(\beta)}{2}, ~~ N_{\overline{1}}=\frac{wt(\alpha \boxminus \beta)+wt(\beta)-wt(\alpha)}{2}.
$$
\end{lemma}

\par

Our second skill is to transfer $\sum\limits_{0\leq i<T} (s_i-s_{i+\tau}) 2^i, ~~~ s_i-s_{i+\tau}\in \{\overline{1},0,1\}$ in Eq.(\ref{S-S}) to a binary expansion with coefficients in $\{0,1\}$.
We need to determine a relationship between the number of non-zeros in $s_i-s_{i+\tau}\in \{\overline{1},0,1\}$ for $0\leq i<T$ and the number of ones in the coefficients of the binary expansion.
We describe the relationship in general case as follows.

\begin{lemma}\label{lemma-transfer}
For any $T$-dimensional vectors $\gamma= (c_0,c_1,\ldots,c_{T-1})\in \{\overline{1},0,1\}^T$
with $N_{\overline{1}}=|\{i : c_i=\overline{1}, 0\leq i<T\}|$
and $\delta=(d_0,d_1,\ldots,d_{T-1})\in \{0,1\}^T$. If
$$
\sum\limits_{0\leq i<T} c_i2^{i} = \sum\limits_{0\leq i<T} d_i2^{i},
$$
we have
$$
wt(\gamma)-N_{\overline{1}}\leq wt(\delta) \leq T-N_{\overline{1}}.
$$
\end{lemma}
Proof. For each block of the form
$\mu=\overline{1}0\cdots 01$ of length $k+2$, i.e., there are $k(\geq 0)$ many zeros in the middle, we see that
$\mu$ corresponds to the decimal integer\footnote{To describe the block with elements in $\{\overline{1},0,1\}$ as an integer, we always assume that the least significant digit is read from the left, for example,
$\overline{1}00\overline{1}1$ corresponds the integer $(-1)\times 2^0+0\times 2^1+0\times 2^2+ (-1)\times 2^3+1\times 2^4=7$.} $(-1)+2^{k+1}=1+2+2^2+\ldots+2^{k}$, so we get a binary block $\nu=1\cdots 10$ of length $k+2$.
We derive
\begin{equation}\label{block-wt}
wt(\overbrace{1\cdots 10}^{k+2})=k+1=2+(k-1)=wt(\overbrace{\overline{1}0\cdots 01}^{k+2})+(k-1).
\end{equation}

Now we begin from the most right bit of $\gamma$ to consider the block of the form $\overline{1}0\cdots 01$ of length $k_1+2$.
 When we replace $\overline{1}0\cdots 01$ by $1\cdots 10$ in $\gamma$ to get $\delta_1$, we see that by Eq.(\ref{block-wt})
$$
wt(\delta_1)=wt(\gamma)+k_1-1.
$$

Then in a similar way, we replace the block of the form $\overline{1}0\cdots 01$ of length $k_2+2$ in $\delta_1$ (from the most right bit)
by $1\cdots 10$ to get $\delta_2$, and we derive
$$
wt(\delta_2)=wt(\delta_1)+k_2-1= wt(\gamma)+(k_1-1)+(k_2-1).
$$

Along the same line, after finite steps, we get $\delta\in \{0,1\}^T$ from $\gamma\in \{\overline{1},0,1\}^T$.
Indeed, we only need to deal with $N_{\overline{1}}$ many steps, since each step helps us to cancel one $\overline{1}$.
So finally we obtain
\begin{equation}\label{total-wt}
 wt(\delta)= wt(\gamma)+\sum\limits_{1\leq j\leq N_{\overline{1}}}(k_j-1).
\end{equation}

It is clear that $0\leq k_j\leq T-wt(\gamma)$, which is the number of zeros in $\gamma$, for $1\leq j\leq N_{\overline{1}}$. We also find that each $0$ in $\gamma$ appears in the block $\overline{1}0\cdots 01$ at most
once\footnote{Some zeros in $\gamma$ might not appear in the block $\overline{1}0\cdots 01$, and those zeros produced in each step will not appear in later blocks $\overline{1}0\cdots 01$.}. Then we get
$$
0\leq \sum\limits_{1\leq j\leq N_{\overline{1}}}k_j\leq T-wt(\gamma),
$$
from which we complete the proof by Eq.(\ref{total-wt}).  \qed\\

For example, let us transfer $\gamma=1\overline{1}00\overline{1}0\overline{1}1\overline{1}0100$ to $\delta$ over $\{0,1\}$ of the same length. We begin from the right of $\gamma$ and firstly consider the block
$\overline{1}01$ in $1\overline{1}00\overline{1}0\overline{1}1\uwave{\overline{1}01}00$,
we replace $\overline{1}01$ by $110$ (since $(-1)+2^{2}=1+2$) and get $1\overline{1}00\overline{1}0\overline{1}111000$. Now let us consider the second block
$\overline{1}1$ in $1\overline{1}00\overline{1}0\uwave{\overline{1}1}11000$, we replace $\overline{1}1$ by $10$ and get $1\overline{1}00\overline{1}01011000$. Then the third block $\overline{1}01$ comes to us
from $1\overline{1}00\uwave{\overline{1}01}011000$, in which after we replace $\overline{1}01$ by $110$,
we finally deal with the block $\overline{1}001$ and hence get $\delta=1111010011000$. Clearly, we have
$$
wt(\delta)=wt(\gamma)+0+(-1)+0+1=wt(\gamma)=7,
$$
in which we apply Eq.(\ref{total-wt}) for four related blocks.

\subsection{Proof of Theorem \ref{thm-AC}}

\indent

From now on, let $\mathcal{M}=(m_i)_{i\geq 0}$ be a binary $m$-sequence of period $T=2^n-1$ for some $n\in \mathbb{N}$.
The $\tau$-shift of $\mathcal{M}$ is denoted by $\mathcal{M}^{(\tau)}=(m_{i+\tau})_{i\geq 0}$.
We view the first period of $\mathcal{M}$ as a $T$-dimensional vector $\omega$ in $\{0,1\}^{T}$, i.e., $\omega=(m_0,m_1,\ldots,m_{T-1})$.
Similarly, we use $\omega^{(\tau)}=(m_{\tau},m_{\tau+1},\ldots,m_{\tau+T-1})\in \{0,1\}^{T}$ for the first period of $\mathcal{M}^{(\tau)}$. 
We need the following 2 steps.  

\textbf{Step 1.} We determine the weight of $\omega \boxminus \omega^{(\tau)}\in \{\overline{1},0,1\}^T$ for $1\leq \tau<T$ and the distributions of $\overline{1},0,1$.
By the ``shift-and-add property" of $\mathcal{M}$ \cite[Theorem 5.3]{GG}, we see that $\omega \oplus \omega^{(\tau)}$ is the first period of a $\tau'$-shift of $\mathcal{M}$
for some $\tau'$. So by Lemma \ref{lemma-boxminus}, we have
$wt(\omega \boxminus \omega^{(\tau)})=wt(\omega \oplus \omega^{(\tau)})=2^{n-1}$, since there are $2^{n-1}$ many ones in one period of $\mathcal{M}$ and $2^{n-1}-1$ many zeros.

On the other hand, together with $wt(\omega)=wt(\omega^{(\tau)})=2^{n-1}$, by Lemma \ref{lemma-boxminus} again one can easily check that
  there are $2^{n-2}$ many $\overline{1}$'s, $2^{n-2}$ many $1$'s
and $2^{n-1}-1$ many 0's
  in the vector $\omega \boxminus \omega^{(\tau)}$.\\

\textbf{Step 2.} We determine the weight of the (unique) vector $\vartheta^{(\tau)}=(d_0,d_1,\ldots,d_{T-1})\in \{0,1\}^T$
such that $\omega \boxminus \omega^{(\tau)}=(m_0-m_{\tau},m_1-m_{\tau+1},\ldots, m_{T-1}-m_{\tau+T-1})\in \{\overline{1},0,1\}^T$
satisfying with
$$
\left| \sum\limits_{0\leq i<T} (m_i-m_{\tau+i})2^{i} \right|=\sum\limits_{0\leq i<T} d_i2^{i}.
$$
We have directly from Lemma \ref{lemma-transfer} and discussions in Step 1,
\begin{equation}\label{bound}
2^{n-2}\leq wt(\vartheta^{(\tau)})\leq 3\times 2^{n-2}-1.
\end{equation}

So, by Eqs.(\ref{AA+}) or (\ref{AA-}), we use Eq.(\ref{bound}) to get
$$
-(2^{n-1}-1) \leq \mathcal{A}^{A}_{\mathcal{M}}(\tau) \leq  2^{n-1}-1,
$$
which completes the proof of Theorem \ref{thm-AC}. \qed

\section{Final remarks and conjecture}

\indent

From Theorem \ref{thm-AC}, we find that the upper bound is about half of the period, it is rather large and achievable on the basis of the following experimental evidence. So it seems that $m$-sequences are far away from looking random from the point of view of the arithmetic autocorrelation.

We run program C to check some examples. For the $7$-periodic $m$-sequence $\mathcal{M}=(0011101)$ (in this case $n=3$),
we have
$$
\mathcal{A}^{A}_{\mathcal{M}}(1)=-1, \mathcal{A}^{A}_{\mathcal{M}}(2)=1,
\mathcal{A}^{A}_{\mathcal{M}}(3)=3, \mathcal{A}^{A}_{\mathcal{M}}(4)=-3,
\mathcal{A}^{A}_{\mathcal{M}}(5)=-1,\mathcal{A}^{A}_{\mathcal{M}}(6)=1.
$$
For $n=4$ and the $15$-periodic $m$-sequence $\mathcal{M}$ produced by the minimal polynomial $X^4+X+1$ or $X^4+X^3+1$, we have
$$
\mathcal{A}^{A}_{\mathcal{M}}(\tau)\in \{\pm 1, \pm 3, \pm 7 \}, ~~~\mathrm{where}~~~ 1\leq \tau< 15.
$$
We also check all $m$-sequences (of period $2^n-1$) for $5\leq n\leq 9$, see Table \ref{table-m-sequence}.
In fact, for larger $n$, for example, we check $m$-sequence $\mathcal{M}$ produced by the minimal polynomial $X^{10}+X^3+1$
or $X^{10}+X^4+X^3+X+1$ for $n=10$, by $X^{11}+X^5+X^3+X+1$ or $X^{11}+X^5+X^3+X^2+1$  for $n=11$, by  $X^{12}+X^6+X^4+X+1$ or
 $X^{12}+X^6+X^5+X^3+1$ for $n=12$ and by $X^{13}+X^5+X^4+X^2+1$ or $X^{13}+X^6+X^4+X+1$ for $n=13$.
 All experimental results are entirely consistent with Theorem  \ref{thm-AC}.

\begin{table}[ht]
\centering
\begin{tabular}{ccc}
\hline\noalign{\smallskip}
 $n$            & How many primitive           & $\mathcal{A}^{A}_{\mathcal{M}}(\tau)\in$     \\
                &  polynomials of degree $n$   &   \\
\noalign{\smallskip} \hline \noalign{\smallskip}
5               &      $6$                     &   $\{\pm 1, \pm 3, \pm 7, \pm 15\}$           \\
6               &      $6$                     &   $\{\pm 1, \pm 3, \pm 7, \pm 15,\pm 31\}$    \\
7               &      $18$                   &   $\{\pm 1, \pm 3, \pm 7, \pm 15,\pm 31,\pm 63\}$    \\
8               &      $16$                    &    $\{\pm 1, \pm 3, \pm 7, \pm 15,\pm 31,\pm 63,\pm 127\}$           \\
9               &      $48$                    &    $\{\pm 1, \pm 3, \pm 7, \pm 15,\pm 31,\pm 63,\pm 127,\pm255\}$           \\
\hline
\end{tabular}
\caption{$m$-sequence $\mathcal{M}$ produced by the primitive polynomial}\label{table-m-sequence}
\end{table}

From the numerical experiments, we find an interesting phenomenon (on the values of the arithmetic autocorrelation) and so
conjecture below:

\begin{conjecture}
Let $\mathcal{M}$ be a binary $m$-sequence of period $2^n-1$ for some $n\in \mathbb{N}$. Then,

(1). the arithmetic autocorrelation of $\mathcal{M}$ satisfies
$$
\mathcal{A}^{A}_{\mathcal{M}}(\tau)\in \{\pm (2^{k}-1) : 1\leq k<n\}, ~~~\mathrm{where}~~~ 1\leq \tau< 2^n-1.
$$
(2). there are $2^{n-k}$ many $1\leq \tau< 2^{n}-1$ such that $|\mathcal{A}^{A}_{\mathcal{M}}(\tau)|= 2^{k}-1$ for $1\leq k<n$.
\end{conjecture}

Although we only focus on binary $m$-sequences, it also works for all binary sequences (such as GMW-sequences) of period $2^n-1$ with ideal (classical) autocorrelation \cite{NGGLG1998}.
In order to sharpen the bounds on $\mathcal{A}^{A}_{\mathcal{S}}(\tau)$,
it is interesting to estimate the value of
$wt(\vartheta^{(\tau)})$ in Step 2 (or $\sum_{1\leq j\leq N_{\overline{1}}}k_j$ in the proof of Lemma \ref{lemma-transfer}) for the considered binary sequence $\mathcal{S}$.

Motivated by the works on the Legendre sequences
and the Sidelnikov-Lempel-Cohn-Eastman sequences by Hofer, Merai and Winterhof \cite{HMW2017,HW2017}, we leave some open problems about
 the arithmetic autocorrelation for other (generalized) cyclotomic generators, such as the Ding-Helleseth-Lam sequences, the Hall's sextic residue sequences,
 the Jacobi(or two-prime) sequences \cite{CDR1998} and the Fermat quotient sequences \cite{C2014,CD2013,WXCK2019} for future study.

Finally we note that, as a byproduct, we can show along the way in Sect.2 a relation between the arithmetic autocorrelation $\mathcal{A}^{A}_{\mathcal{S}}(\tau)$ and
the classical autocorrelation $\mathcal{A}_{\mathcal{S}}(\tau)$ of binary $\mathcal{S}$ of period $T$:
$$
\left| \mathcal{A}^{A}_{\mathcal{S}}(\tau)\right|\leq  \frac{T+\mathcal{A}_{\mathcal{S}}(\tau)}{2},  ~~~\mathrm{where}~~~ 0\leq \tau< T.
$$

\section*{Acknowledgments}

Z. Chen and C. Wu were partially supported by the National Natural Science
Foundation of China under grant No.~61772292, and by the Provincial Natural Science
Foundation of Fujian, China under grant No.~2020J01905.
Z. Niu and Y. Sang  were partially supported by the National Key Research and Development Program
of China under grant No.~2018YFB0704400, and by the Key Laboratory of Applied Mathematics of Fujian Province University (Putian University) under grant No.~SX202102.

\end{document}